\documentclass[twocolumn]{article}
\title{Towards  Quantum Mechanism of Sensory Transduction 
\\
\strut
\\
\normalsize 
Oleg A.KHRUSTALEV
\\
 N.N.Bogoliubov's Institute of Theoretical Microphysics,
Moscow State University, \\
Moscow, 119899,  Russia
\\ {\it e-mail: khrust@bog.msu.ru}
\\
and
\\
Olga D.TIMOFEEVSKAYA
\\
Physics Deparment, Moscow State 
University, \\ Moscow, 119899,  Russia 
\\ {\it e-mail: olga@goa.bog.msu.ru}}

\date{ \ \ \  }
\author{ \ \ \  }

\begin{document}

\maketitle
\begin{center}
  
{\bf ABSTRACT}

\end{center}

In this communication we  consider wave equations describing
the classical excitations in microtubules. 
We found that there exists  double-periodic solution for these 
equations. Classical solutions form background of quantum 
excitations. Quantum excitations are considered by Bogoliubov's group
variables method. As a result these contain no zero-modes. 

\vskip 0.8 truecm


\vskip 1. truecm

\begin{center}
  
{\bf 1. INTRODUCTION}

\end{center}

Among the various structures of cytoskeleton  microtubules
appear to be  prominent ones (and almost unknown for 
nonbiologists).  Microtubules represent hollow cylinders formed by
protofilamets aligned along their axes. The cylindrical walls of
microtubules usually are assemblies of 13 longitudinal protofilaments,
each of which is the series of subunit proteins --- tubulin dimers.
Each dimer has an electric dipole. Thus from the
physical point of view microtubules may be described as oriented
 assemblies of dipoles.

The hypothesis that microtubules are actively involved in 
reception and transduction of sensory information was been proposed in
the twenties of the past century [1].

Frohlich [2] developed the general theory
of  biological structures which allows  nonlinear dipole excitations
among substructures. Frohlich theory asserts that the coherent
oscillations and the long-range order arise in biological structures.

The dimers form  the hexagonal
structures on the outer well of microtubules. Hameroff
[3] connected their properties with possibility of the cellular automat. 
The states of this automat are changed as a result of Frohlich dipole 
oscillations. This array was the smallest classical computer.

Recently it was assumed [4] that  the kink-like
 excitations may propagate along the microtubules(with velocity
about 1 m/s) and that  creation and  detection  of these excitations
develop into the communication system of living organisms.

However there was serious objection against this model. It was connected
with charge transport along  microtubules. In particular Tegmark
[5] has showed that this mechanism could not lead to the
spread of excitations.

In this communication we study nonlinear wave equations describing
 microtubule excitations. We found that there exists  double
-periodic solutions for these nonlinear wave equations.

 These solutions are connected with charge transport
along the microtubules
and define the structure of quantum excitations arising on classical
background.
There is no problem with coherence in this case.

The problem of quatization of nonlinear wave equations in the
neighborhood
of classical field, here classical double-periodic solution,
is considered in this paper by Bogoliubov's group
variables method [6]. One of very important achievement
of  Bogoliubov's
method is absence of so named zero modes arising in straight methods
of quantization on classical background. Thus this approach gives us
 possibility to avoid  false conclusion about long-range interactions
 in biological systems. The symmetry of the system connected
 with oscillating
 motion of dimers is taking into account precisely. The introduction
 of the new dynamical variables
 which play the role
of symmetry group parameters gives us possibilities to
 calculate the quantum
corrections as perturbation to classical solution and to take account the
conservation laws.

We propose the scheme of quantization  and receive
the quantum excitations in the neighborhood classical solutions. These
results leads to the assumption that  quantum communication can
play the important role in  sensory and cognition process.
In particular it is possible to separate the structure inside the
mucrotubules such that their  function is similarly to quantum automats.
 The operating speed  for quantum automats as usual is higher than
 the operating speed for corresponding classical automats.
 In addition the acceleration of operation is connected with
 higher organization of microtubule states but not  with new physical
 mechanisms.

The analyze of that possibilities demands the higher than usual
precision in definition of notions for divisions quantum systems into
subsystems and reunification of subsystems into a joint system.
This problem always arises in investigation of quantum scheme communications.
The consideration of these processes is carried out with the help
of conception of conditional density matrix [7]. This
approach could  solve all
problems well known as "quantum paradoxes". Consideration in the terms
of conditional density matrix gives adequate to physical process description.

\vskip 0.8 truecm

\begin{center}

{\bf 2. CLASSICAL FIELD }

{\bf IN MICROTUBULES }

\end{center}

To considerable extend
the physical foundation of Penrose's and  Hammeroff's cognition 
[8],[9] theory is based on the hypothesis that there exist solitons 
in microtubules such that "quantum computation" processes 
are  connected  with them.
It is very essential to the understanding of similar processes
that although the generation of solitons is absolutely quantum
effect the superposition of an abundance of quantum modes
hides most of pure quantum soliton properties. As a result
in the first place the soliton should be considered as classical
object connected with the process of a quantum cognition.

In this paper we want to show that there is a possibility
of generation  a different state in
 microtubules on the classical level. This state could forms a
 background for an important "quantum cognition" processes.

 The description of quantum cognition processes is based on the 
conception the nature of 
biological order after Frohlich [2]. Frohlich argued 
that an ensemble of strongly interacting electrical dipoles that are 
capable of high-frequency oscillations under the influences of the 
external electric field may form the a metastable state 
characterized by long-range correlations. 

Let the field $\phi (x)$ be a vector projection on the microtubule 
axis for microtubule dimer deflection from equilibrium.
Here $x$ is a coordinate along the microtubule axis. Moreover
we suppose that the time evolution of the field is described by
nonlinear Klein-Gordon equation.
$$   {{\partial}^{2}{\phi}(x,t) \over {\partial}t^{2}}
     - {{\partial}^{2}{\phi}(x,t) \over {\partial}x^{2}} \quad + 
$$
$$
         m^{2}{\phi}(x,t) 
     +  {\epsilon}{\phi}^{3}(x,t)
          = 0,  
$$
$$       0 \leq x  \leq 2{\pi}.                          \eqno{(1)}
$$
Here the electrical dimers field [4] 
influence on the orientation of the field $\phi$ is not taken
 into account. Nevertheless these effects arise during successive
description for micrortubules processes on the basis of
nonstationary polaron theory {\bf [11]}.

It is known two classes of double periodic solutions for
this equation. The first one is traveling wave:
$$   \phi(x,t) \quad = \quad {\phi}(x - vt). 
 \eqno{(2)}
$$
In this case the wave equation is Duffing equation.
The periodic solutions are elliptic Jacobi functions.

The second class of doubly periodic solutions consists of function 
in the standing wave form:
$$  {\phi}(x,t)  =  $$
 $$  \sum_{k=1}^{\infty} \sum_{l=1}^{\infty}
     C_{kl}\sin(k(x - x_{0}))\sin(l{\omega}(t - t_{0})),
 \eqno{(3)}
$$
where $x_{0}$ and $t_{0}$ are constants determined by boundary and 
initial conditions. The wave equation is a translation-invariant, so 
we can restrict our consideration to  zero $x_{0}$ and 
$t_{0}$.

In this case the solutions were under study in the paper
[11] provided that $\epsilon$ is small.
 
The possibility to obtain an uniform expansion, using standard 
asymptotic methods, depends on value of the frequencies in the zero 
approximation. 

If $\epsilon = 0$, then wave equation is linear and 
 has periodic solutions with frequencies in time ${\Omega}_{l} = 
\sqrt{l^{2} + m^{2}}$. There are two fundamentally different cases. 
If the ratio ${\Omega}_{j} / {\Omega}_{k}$ is irrational for all 
${\it j}$ and ${\it k}$ then it is a non-resonant case. The periodic 
asymptotic in non-resonant case are well-known.

The resonance case, when there exist two frequencies ${\Omega}_{j}$, 
whose relation is rational, is more difficult. The Krylov-Bogoliubov 
method  can be used to find periodic solutions only to a few leading 
orders in $\epsilon$.

Without loss of generality it can be assumed that under resonance
conditions $m = 0$. 
We introduce the new time ${\tilde t} = {\omega}t$ and look for a 
doubly-periodic solution in the form:
$$ {\phi}(x, {\tilde t}, \epsilon)  = 
    \sum_{n=0}^{\infty}{\epsilon}^{n}{\phi}_{n}(x, {\tilde t}),
 \eqno{(4)}
$$ 
$$   \omega \quad = \quad {\omega}(\epsilon) 
        \quad = \quad
        1  +  \sum_{n=1}^{\infty}{\epsilon}^{n}{\omega}_{n}.
 \eqno{(5)}
$$ 

Expanding the wave equation in  power series in $\epsilon$ gives a
sequence of equations for ${\phi}_{n}$. Two leading  equations 
are:  $$   {{\partial}^{2}{\phi}_{0}(x,{\tilde t}) \over 
      {\partial}{\tilde t}^{2}}
      - {{\partial}^{2}{\phi}_{0}(x,{\tilde t}) \over {\partial}x^{2}}
           \quad = \quad 0,  \eqno{(6)}
$$
$$   {{\partial}^{2}{\phi}_{1}(x,{\tilde t}) \over 
      {\partial}{\tilde t}^{2}}
     \quad - \quad
     {{\partial}^{2}{\phi}_{1}(x,{\tilde t}) \over {\partial}x^{2}}
           \quad =$$
$$  
     2{\omega}_{1}{{\partial}^{2}{\phi}_{0}(x,{\tilde t}) \over 
      {\partial}{\tilde t}^{2}}
          \quad + \quad 
      {{\phi}_{0}}^{3}(x,{\tilde t}).  \eqno{(7)}
$$
The general solution of the zero approximation is
$$   {\phi}_{0}(x,{\tilde t})
      =  
       \sum_{n=1}^{\infty}a_{n}\sin(nx)\sin(n{\tilde t})
   \eqno{(8)}
$$
with arbitrary $a_{n}$. We have to find coefficients $a_{n}$ so that 
the function ${\phi}_{1}(x,{\tilde t})$ is a double-periodic:
$$   {\phi}_{1}(x,{\tilde t})
      = 
       \sum_{k=1}^{\infty}\sum_{l=1}^{\infty}
      b_{kl}\sin(kx)\sin(l{\tilde t}).  \eqno{(9)}
$$
Whenever  quantum phenomena for system with essentially
nonlinear interaction are investigated it must be borne in mind
that quantization develops on the classical field background.
In other words the ground state of the system is no longer
vacuum, i.e. the state without any quantum excitation, but
another state. New "dressed vacuum" contains so many quantum
excitations that it is easier to describe its properties
on classical language. Namely this property of the ground state
is principal for Frolich approach.
The fundamentals for general theory of similar phenomena
were found by Bogoliubov in 1950 [6].

We present the model of quantum information transport inside
microtubules in terms of representation about nonstationary
polaron [11].
Let  ${\Psi}(x)$ be the field connected with  dimer distribution
for microtubules; then it is possible to present the action
of the system in the form
$$  S \quad = \quad 
  {1 \over 2}{\int}d{\tau} ({\phi}_{t}^{2} - {\phi}_{x}^{2} - 
       m^{2}{\phi}^{2}) \quad +  $$ 
$$            g^{2}{\int}d{\tau} ({\Psi}^{*}_{t}{\Psi}_{t} - 
   {\Psi}^{*}_{x}{\Psi}_{x} - M^{2}{\Psi}^{*}{\Psi}) 
         \quad - $$
$$ 
   g{\int}d{\tau}{\Psi}^{*}{\Psi}{\phi}^{2},    \eqno{(10)}
$$
The translation invariance of the action results in two integrals of 
motion, the energy and the momentum. We obtain the expressions for 
the integrals of motion
$$  H \quad = \quad 
   {1 \over 2}{\int}d{\lambda} ({\hat p}^{2} + {\hat q}_{\lambda}^{2} 
       + m^{2}{\hat q}^{2}) \quad + \quad 
$$
$$               g^{2}{\int}d{\lambda} 
    ({\Psi}^{*}_{t}{\Psi}_{t} - {\Psi}^{*}_{\lambda}{\Psi}_{x} - 
   M^{2}{\Psi}^{*}{\Psi})  \quad - $$
$$ 
        g{\int}d{\lambda}{\Psi}^{*}{\Psi}{\hat q}^{2},  \eqno{(11)}
$$
$$  P  =  {\int}d{\lambda}{\hat p}{\hat q}_{\lambda}
                +     g^{2}{\int}d{\lambda}
    ({\Psi}^{*}_{n}{\Psi}_{\lambda}  + 
            {\Psi}^{*}_{\lambda}{\Psi}_{n}).    \eqno{(12)}
$$
Inequality  $g >> 1$ clearly shows the energy relation between
different subsystems: if the interaction is neglected; then monomers
energy is rather more than the energy associated with the field $\phi$.

To define quantized field $\phi$ it is necessary to define the operators
$\hat q$, $\hat p$ such that the canonical commutation relation satisfies

$$  [{\hat q}(x,t), {\hat p}(x^{'},t)] 
             \quad = \quad 
          i{\delta}(x - x^{'}).     \eqno{(13)}
$$
For our purposes the next representation is best suitable
$$   {\hat q}(x) \quad = \quad
    {1 \over \sqrt{2}}({\phi}(x)  + 
       i{{\delta} \over {\delta}{\phi}_{n}(x)}),  \eqno{(14)}
$$

$$   {\hat p}(x) \quad = \quad
    {1 \over \sqrt{2}}({\phi}_{n}(x)  - 
       i{{\delta} \over {\delta}{\phi}(x)}),   \eqno{(15)}
$$
It is supposed that the functions 
 ${\phi}(x)$, ${\phi}_{n}(x)$ are independent here.

\vskip 0.8 truecm

\begin{center}

{\bf 3. PERTURBATION THEORY}

\end{center}

After separating large components from these functions
$$   {\phi}(x) \quad = \quad gv(x^{'}) 
    \quad + \quad u(x^{'}),     \eqno{(16)}
$$
$$  {\phi}_{n}(x) \quad = \quad gv_{n}(x^{'})
     \quad + \quad u_{n}(x^{'}),    \eqno{(17)}
$$
 one gets that the operators $\hat q$, $\hat p$ are transformed analogously.
 The terms have different powers of  $g$:
$$   {\hat q}  = 
   gF(x^{'}) + {\hat Q}(x^{'}) + {1 \over g}{\hat A}(x^{'}),
 \eqno{(18)}
$$
$$   {\hat p} =  
   gF_{n}(x^{'}) + {\hat P}(x^{'}) + {1 \over g}{\hat A}_{n}(x^{'}).
 \eqno{(19)}
$$
Hamiltonian has a similar structure:
$$   {\hat H}  = 
      g^{2}H_{-2} + g{\hat H}_{-1} + {\hat H}_{0}
     + {1 \over g}H_{1}. \eqno{(20)}                                          
$$
The first term is c -- number.The operator
 ${\hat H}_{-1}$  is linear form of $\hat Q$ ¨ $\hat P$ and the operator
 ${\hat H}_{0}$  is bilinear form of  
 $\hat q$, $\hat p$, ${\Psi}$, ${\Psi}_{n}$, ${\Psi}^{*}$,
$\Psi^{*}_{n}$.
The linear powers of
 $\hat Q$ and $\hat P$ arise before quadratic ones. This circumstance
inhibits the realization of regular perturbation theory. For 
elimination of these difficulties we put ${\hat H}_{-1}$ equal zero. It
is possible only in the case when  function 
  $v(x,t)$ satisfies the equation
$$  v_{tt}(x,t) \quad - \quad v_{xx}(x,t) \quad +$$
$$ m^{2}v(x,t) \quad + \quad W(x,t) \quad = \quad 0,  \eqno{(21)}
$$
where $W(x,t)$ is functional of the third order according to 
 $v(x,t)$.

Degree of nonlocality is defined by effective dimension of microtubule
dimer. To a first approximation when  dimer is considered as point
 particle  the equation for $v$  becomes nonlinear Klein-Gordon equation
which was analyzed earlier.

\vskip 0.8 truecm

\begin{center}
 
 {\bf 4. CONDITIONAL DENSITY MATRIX }

\end{center}

 After transformations made above we are close to construction
of realistic model of quantum excitations in microtubules. Nevertheless
 the description how these excitations work
needs the additional efforts.
The famous Penrose and Hammeroff theory  tries to overcome the
arising difficulties. The proposed theory is very ingenious. However
we want to note that in our opinion the authors are not completely right.

 Really  quantum description of quantum brain function is a
problem of quantum communication theory. Its solution  needs
creation of systematic methods for description of subsystem
 unification and system separation into subsystems. Although
this problem was resolved in 1927 by von Neumann the present
researchers as rule do not use the advantages of his approach and
commonly  methods used for  
system and subsystems description are not quite clear. The density
matrix method could be developed by conditional density matrix
introducing. We suppose that conditional density matrix utilization
 helps to resolve the problems of quantum problem description in
microtubules.
 
The possibility to separate the system $S_{12}$ into subsystems
 $S_{1}$  and  $S_{2}$ naturally leads to following
definition. 

Let an operator 
$\hat F = \hat f(x) \hat P_2 (y) $, where
$\hat P_2$ is a projector on the ${\Psi}_{2}$,correspond to the
 physical variable $F(x,y)$  then 
$$ < \hat F >_{\rho} = Tr_{1,2} (\hat f \hat \rho \hat P_2 )=
 Tr_1 (\hat f Tr_2(\hat \rho \hat P_2 )).$$
The operator $P_{2}$ may be connected with the state of subsystem II  
such that some observable $F_{2}$ has an exact value $f_{2}$.
The operator $Tr_{2}({\hat \rho}{\hat P}_{2})$ is proportional to some 
density matrix of the first system -- ${\tilde \rho}_{I}$. 
The ${\tilde \rho}_{I}$ defines state of system I under condition
 that $F_{2}$ has the 
exact value $f_{2}$. 
After normalization we get an operator
$$ \hat \rho _{1/2} = {Tr_2 (\hat P_2 \hat \rho) \over Tr(\hat P_2 \hat
\rho ) }
$$
It is Conditional Density Matrix for the subsystem 1 when the subsystem
2 is selected in the pure state $\hat \rho _2 =\hat P_2 = \hat \rho _2 ^2$.
 It is the most important case for quantum communications. 

\vskip 0.8 truecm

\begin{center}

 {\bf 5. CONCLUSION}

\end{center}

  In present communication for the first time the successive description
 of interaction of the 
field  ${\phi}(x)$  with tubulin dimers is proposed.
The field  ${\phi}(x)$ is deflection vector projection of dimers on the
 axis of microtubule with finite length.

 In the proposed formalism the new type of space- time microtubule
 organization is found out. This organization is connected
with double-periodic solution ${\phi}(x)$

Quantum description for microtubules processes is made with utilization 
of Bogoliubov's method. This method givesq us possibility to take into 
account the existence of classical component and translation invariant 
quantum excitations simultaneously. 

The processes of unification and separation composite system into
subsystems are basic for information theory. It is proposed
to use  conditional density matrix method for construction 
information quantum transport mechanism in microtubules.

\vskip 1 truecm

\begin{center} 

{\bf REFERENCES}

\end{center}

$[1]$ A.E.Hopkins, J. comp. Neurol., Vol. 41, 1926, pp. 253.

[2] H.Frohlich, "Evidence of bose condensation-like
excitation of coherent modes in biological systems",
Physics Letters, Vol.51A, No.1,1975, pp,21-22.

[3] S.P.Hameroff, R.C.Watt, "Information processing
in microtubules", J.Theor.Biol., Vol.98, 1982, pp.549-561.

[4] M.V.Satarich, J.A.Tuszy\'nski, R.B.Zakula, "Kinklike
excitations as an energy-transfer mechanism in microtubules",
Physical Review E, Vol.48, No.4, 1993, pp.589-597.

[5] M.Tegmark, "Importance of quantum decoherence in brain
processes", Phys.Rev., Vol.{\bf E 61}, No.4, 2000,
pp.4194-4206.

[6] N.N.Bogoliubov, Ukranian Math.Journal,
Vol.2, 1950, pp.2-34.

[7] V.V.Belokurov, O.A.Khrustalev, V.A.Sadovnichy,
O.D.Timofeevskaya,"Systems and subsystems in quantum
communication", Preprint, Moscow State University, 
Moscow, 2001; arXiv:quant-ph/0111164.

[8] S.P.Hameroff, "Quantum computation in brain microtubules?
The Penrose-Hamerof "Orh Or" model of consciousness",
Philosophical Transactions  Royal Society London (A),
Vol.356, 1998, pp.1869-1896.

[9] R.Penrose, The Emperer's New Mind, Oxford Press, Oxford,
U.K., 1989.

[10] O.Khrustalev, S.Vernov, 
 Construction of doubly periodic 
solutions via the Poincare-Lindstedt method in the case
of massless ${\phi}^4$ theory, 
Mathematics and Computers in Simulation, Vol. 57,
 2000, pp. 239-252.

[11] E.Yu.Spirina, O.A.Khrustalev, M.V.Tchichikina,
"Nonstationary polaron", Theor.and Math.Physycs,
Vol.122, No.3, 2000, pp.347-354.

[12] S.R. Hameroff, R. Penrose, Orchestrated reduction of quantum 
coherence in brain microtubules: A model for consciousness, Neural 
Network World", Vol. 5, No. 5, 1995, pp.793-804.

\end{document}